# Secret Image Sharing Using Grayscale Payload Decomposition and Irreversible Image Steganography

**Abstract.** To provide an added security level most of the existing reversible as well as irreversible image steganography schemes emphasize on encrypting the secret image (payload) before embedding it to the cover image. The complexity of encryption for a large payload where the embedding algorithm itself is complex may adversely affect the steganographic system. Schemes that can induce same level of distortion, as any standard encryption technique with lower computational complexity, can improve the performance of stego systems. In this paper we propose a secure secret image sharing scheme, which bears minimal computational complexity. The proposed scheme, as a replacement for encryption, diversifies the payload into different matrices which are embedded into carrier image (cover image) using bit X-OR operation. A payload is a grayscale image which is divided into frequency matrix, error matrix, and sign matrix. The frequency matrix is scaled down using a mapping algorithm to produce Down Scaled Frequency (DSF) matrix. The DSF matrix, error matrix, and sign matrix are then embedded in different cover images using bit X-OR operation between the bit planes of the matrices and respective cover images. Analysis of the proposed scheme shows that it effectively camouflages the payload with minimum computation time.

**Keywords:** DSF matrix, error matrix, sign matrix, bit plane

## 1 Introduction

Image steganography is the science of concealing information and messages within an image using some embedding algorithm. In modern days communication security is the fundamental requirement even though accomplishment of comprehensive security is a superlative state of affairs. The concept of ''What You See Is What You Get (WYSIWYG)'' associated with printing capabilities of a printing machine for an image, is now a fallacy [1]. Ordinary image is no longer a regular image. It could be a mask over a secret message. In image Steganography, the mask used to hide any message is a color or grayscale image.

Two extensive approaches used in image steganography are reversible image steganography and irreversible image steganography. Message to be embedded within an image is known as the payload. After embedding the payload in image (more specifically cover image) the resulting image called the stego image is sent to the authorized recipient, where the payload is recovered. Reversible image steganography derives its name from the mode of recovery of the payload wherein the recovered cover image is noiseless. On the contrary irreversible image steganography strives to achieve high capacity embedding without giving much emphasis on the carrier recovered during the extraction process. Reversible image steganography facilitates the easy detection of any alteration in the stego image whereas irreversible image steganography achieve high embedding capacity. Motivation for modern day's image steganography techniques comes from the fact that no existing method is self-sufficient. Increasing the embedded information could cause easy detection by an attacker, whereas enhancing the security could increase the computation and communication overhead due to the fact that one secret message will take several transmissions. A trade-off between embedding capacity and the level of security needs a strong base for a security measure. The proposed scheme tries to achieve greater level of security with minimum computation time.

The failure to cease the suspicion of any hidden data in an image is the reason for breakdown of any steganography system. The fundamental requirement for any secure steganography scheme is the ability to cover the hidden message in the cover image. A system is considered to be secure if a snooper cannot distinguish between cover image and the stego image. There are different measures for steganographic security. The most common measure is called detectability of a stego system. Detectability is defined as the relative entropy between the probability distribution of cover image and the stego image. Any steganography system is called $\varepsilon$-secure if the relative entropy of the system is at most $\varepsilon$ [1]. A steganography scheme is said to be perfectly secure if detectability is zero. Reduction in detectability means reduced embedding capacity. Any image steganography scheme should optimize the embedding capacity to achieve minimum possible detectability taking into account the computational overhead. The maximum number of bits that can be embedded in a cover image and recovered from the stego image without violating the undetectability constraints is known as the steganographic capacity. The maximum steganographic capacity that an existent reversible steganographic scheme can achieve is approximately 3 bpp [2]. Perceptual consistency and robustness are the most desirable attributes of any steganographic system. The capacity of the steganographic system is comprehensively enhanced by irreversible models of image steganography. Peak signal to noise ratio can be used as a measure of steganographic security where the embedding capacity of a model is outsized. Increasing secret information embedding capacity would mean straightforward steganalysis and detection of the hidden information. State of art steganalysis attempt to overpower any steganography scheme. Biggest challenge of a steganography scheme is to outsmart all steganalysis schemes [1]. The science of detecting the concealed message in a cover image is steganalysis. The battle between steganography and steganalysis is getting on since the evolution of the science of steganography. There are several ways in which steganalysis can wreck the structure of steganography. The most common methods are inspection of the inner structure of LSBs, Histogram analysis, feature vector analysis et cetera [1]. Primary goal of any image steganography scheme is to achieve high level of security with high capacity embedding, reduced noise, and minimum computation time.

## 2 Related Work

Many schemes have been proposed in reversible and irreversible image steganography. Major work that is going on in image steganography attempts to frame steganographic structure that effectively hides the payload. Additional security is provided by encrypting the payload with encryption algorithms, this may however increase the complexity of the steganographic scheme two folds.

Chih-Chiang Lee et al. proposed an adaptive lossless image steganography scheme, which embeds variable length secret information into fixed sized blocks of cover image. Amount of information embedded in each block depends on the complexity of the cover image [3]. This method is an improvement over Alattar's scheme based on generalized difference expansion [3], where N successive pixels of cover image are taken to embed N-1 bits of secret information. The improvement is achieved in embedding capacity by the use of a number of non overlapping blocks of size $m \times n$ in place of N-1 difference values to hide the secret information as proposed by Alattar [3]. If the secret information under consideration is a grayscale image of size $a \times b$ and average number of bits embedded in each block of size $m \times n$ is four, then total number of such blocks required is $\frac{(a \times b \times 8)}{4}$. Hence the size of the required cover image is $\frac{\{(m \times n) \times (a \times b \times 8)\}}{4}$. In this scheme secret information is embedded into the centralized difference values without encryption. If a small sized grayscale image is to be embedded using state of art encryption techniques to add another layer of security, enormous computation time may reduce the efficiency of the method.

Reversible image hiding scheme proposed in [4] embeds secret information by shifting the histogram of residual image computed as the difference of the original cover image and the corresponding predictive pixel, called the basic pixel, obtained using linear prediction technique. This scheme is an improvement over reversible data hiding scheme proposed in [5]. The hiding capacity is increased by employing linear prediction technique on the cover image to form the residual image. A similar approach has been proposed by Lin et al. [6]. The secret information embedding scheme proposed in [7] hides encrypted payload into a hiding tree computed from frequency of absolute error values obtained as the difference of the cover image (host image) and predictive image. Predictive image is generated using median edge detector (MED) predictor. Similar reversible image steganography schemes have been proposed in [8-9], where error values are explored to embed the secret information. Although state of art steganography methods hide the secret information comprehensively, encryption is applied on the secret message to provide an additional layer of security. The methods of image steganography proposed in [4-9] are themselves complex enough; hence additional computational complexity of standard encryption techniques, such as AES or DES, can reduce the efficiency of the steganographic system. LSB replacement is the most commonly used irreversible steganographic method. In this method, only the LSB plane of the cover image is replaced with any of the bit plane of the grayscale payload. As a result, the change in bit structure of the original cover image is inevitable, and thus the existence of hidden payload even at a low embedding rate is possible with the use of some existing steganalytic algorithms, such as the Chi-squared attack [10], regular or singular groups (RS) analysis [11] and sample pair analysis [12]. LSB matching (LSBM) is an improvement over LSB replacement. A particular pixel intensity of the cover image is either incremented or decremented by 1 if there is a match between the secret bit and the LSB of that pixel. Several steganalytic algorithms [13-16] have been proposed to analyze the stego image encoded using LSBM scheme. Encryption techniques are used to foil these steganalytic algorithms.

The proposed method diffuses the payload across different matrices to achieve the same visual distortion as any encryption technique. As the payload is distributed into different matrices any intruder has to retrieve all matrices to re-compute the payload accurately. Any image steganography scheme should eliminate the suspicion of the existence of any hidden information within the cover image. What we propose is another level of security to get around the steganalysis scheme which may foil the image steganography used to embed the secret information, without incorporating substantial computational overhead. The proposed scheme achieves non linear polynomial computational growth and better computation time.

## 3 Proposed Scheme

In the proposed method we intend to introduce distortion in the distributed payload with minimum computation time, which appears to be natural distortion caused by the noise in the communication channel. The proposed scheme distributes the payload into three matrices; DSF matrix, error matrix and sign matrix

All three matrices represent distorted form of original payload. Proposed scheme is illustrated in Fig. 1. Down Scaled Frequency matrix generator counts and replaces the grayscales in the payload with the corresponding counts called the frequency. These frequencies are then mapped using a mapping function to produce the final DSF matrix. Sign matrix is a matrix of zeros and ones. Sign matrix generator places a zero in the matrix whenever corresponding grayscale in the payload is greater than or equal to the corresponding Down Scaled Frequency in the DSF matrix. Difference between DSF matrix and the payload called the error matrix is generated by the error matrix generator. Stego system encoder embed these three matrices, having same dimension as the payload, into the least significant bit plane of the cover image using bit X-OR operation with the second least significant bit plane of the cover image.

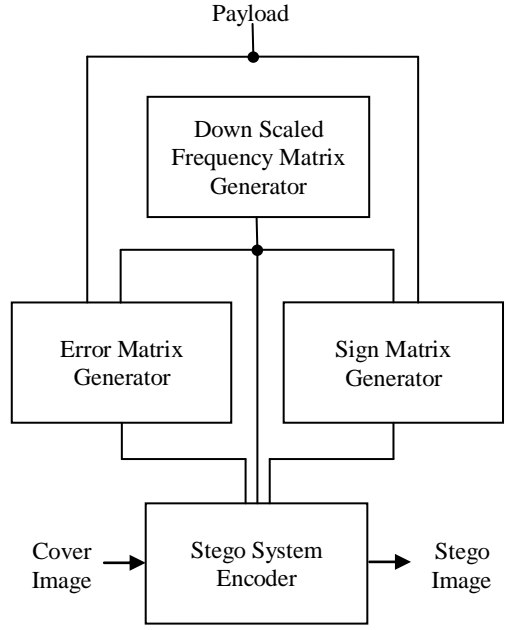

**Fig. 1.** Payload decomposition into frequency, error and sign matrices; stego-image generation.

### 3.1 Matrix Computation

Let us define the payload as

$$G = \left[ g_{ij} \right]_{m \times n} \quad (1)$$

where, m×n is the size of the payload and $g_{ij}$ is the grayscale in the $i^{th}$ row and $j^{th}$ column of the payload. Each element of the DSF matrix is the Down Scaled Frequency value of frequency matrix computed from the payload G. The frequency matrix is defined as

$$F = \left[ f_{ij} \right]_{m \times n} \text{ where } 1 \leq f_{ij} = count\left(g_{ij}\right) \leq m \times n \ \forall g_{ij} \in G \quad (2)$$

Where, *count* ($g_{ij}$) is the number of times $g_{ij}$ appears in the payload. We define a mapping parameter $\mu$ in equation (3) to map the frequency matrix into DSF matrix.

$$\mu = 2^\eta, \eta = 1,2,3,4,\dots 8 \quad (3)$$

We define three more parameters upper bound (*ub*), lower bound (*lb*) and *level* as

$$ub = \mu + k \times \mu, \text{ where } k = 0,1,2,\dots m \times n/\mu \quad (4)$$

$$lb = ub - \mu, \text{ where } k = 0,1,2,\dots m \times n/\mu \quad (5)$$

$$level = k + 1, \text{ where } k = 0,1,2,\dots m \times n/\mu \quad (6)$$

We down scale the frequency of occurrence of a grayscale to avoid the overflow of the frequency value. Down Scaled Frequency matrix is computed using a mapping function denoted as *DSF(.)*.

Down Scaled Frequency matrix is defined as

$$D = \left[ d_{ij} \right]_{m \times n}, \text{ where } d_{ij} = DSF\left(f_{ij}\right) \quad (7)$$

Mapping function *DSF( )* is defined as

$$DSF\left(f_{ij}\right) = \left\lceil f_{ij}/level \right\rceil \quad (8)$$

such that $lb \leq f_{ij} < ub$ and $level = k+1$, where $k = 0,1,2,\dots m \times n/\mu$

To compute the Down Scaled Frequency the frequency value of a grayscale is divided by the corresponding *level* parameter. The greatest common integer of the resulting quotient is the Down Scaled Frequency for that particular grayscale. The value of *level* parameter depends upon the value of *k*, which intern depends on *ub* and *lb* of the frequency of grayscale satisfying the constraint $lb \leq f_{ij} < ub$.

The error matrix E is the difference between G and D.

$$E = \left[ e_{ij} \right]_{m \times n} = \left[ \left| g_{ij} - d_{ij} \right| \right]_{m \times n} \quad (9)$$

Sign matrix S is defined as

$$S = \left[ s_{ij} \right]_{m \times n}, s_{ij} = \begin{cases} 0, g_{ij} \geq d_{ij} \\ 1, g_{ij} < d_{ij} \end{cases} \quad (10)$$

Figure 2(a) shows a 4×4 sample payload with different grayscales. For $\eta = 1$, Table 1 shows values of different parameters corresponding to grayscales in the sample payload. For grayscale $g_{34} = 4$, we have frequency of occurrence $f_{34} = 3$. Every occurrence of grayscale 4 in sample payload is replaced by the corresponding frequency of occurrence to generate the frequency matrix as shown in figure 2(b). If we take $\eta = 1$, then the mapping parameter $\mu$ is 2. The lower bound (*lb*) and upper bound (*ub*) for the frequency value 3 are 4 and 6 respectively as shown in Table 1. From equation (4) we have $k = 2$, for $ub = 6$ and $\mu = 2$. The value of *level* for $k = 2$ is 3 as *level=k+1*. So, the Down Scale Frequency for grayscale 4 is 1, which is computed using the mapping function in equation (8). Figure 2(c) shows the DSF matrix for the sample payload. The error value for the grayscale 4 is 2 from equation (9). The sign matrix for the sample payload is shown in Figure 2(e). All values in sign matrix are zero due to the fact that DSF values in the DSF matrix are less than the corresponding grayscale in the sample payload; moreover if there had been some DSF values in the DSF matrix greater than the corresponding grayscale values then there would have been 1s for those DSF values in the sign matrix.

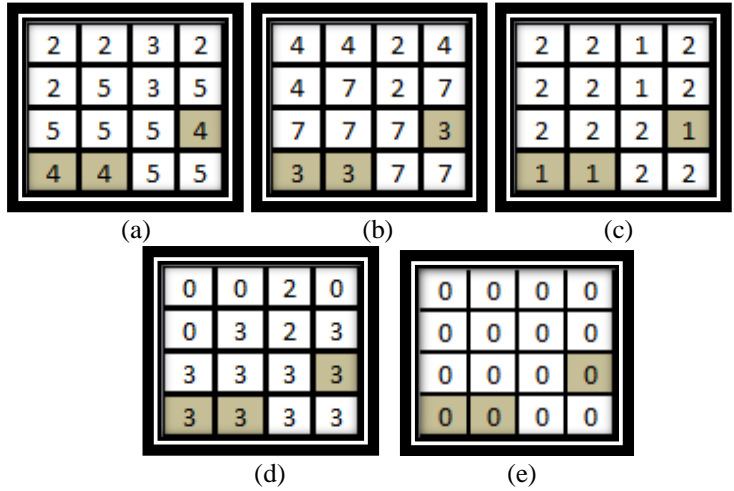

**Fig. 2.** Sample payload and corresponding frequency, DSF and error matrices; (a) sample payload, (b) frequency matrix, (c) DSF matrix, (d) error matrix, (e) sign matrix

**Table 1.** The values of different parameters for frequency and DSF values corresponding to a grayscale in the sample payload

| $\eta$ | $\mu$ | ub | lb | k | level | frequency | DSF |
|---|---|---|---|---|---|---|---|
| 1 | 2 | 4 | 2 | 1 | 2 | 2 | 1 |
| 1 | 2 | 6 | 4 | 2 | 3 | 3 | 1 |
| 1 | 2 | 6 | 4 | 2 | 3 | 4 | 2 |
| 1 | 2 | 8 | 6 | 3 | 4 | 7 | 2 |

The distribution of the payload over DSF and error matrices is a better choice not only because it reduces the computation time but also due to the fact that even if the intruder is familiar with the proposed method of recovery the possibility of prediction of a particular secret payload from either intercepted DSF or error matrix of an anonymous payload is simply not possible. If a payload is encrypted using any standard encryption technique and the intruder is familiar with the encryption used then the intruder only needs to intercept an anonymous encrypted payload that is encrypted with same encryption technique to predict the key used to encrypt a particular payload. Hence it is possible for the intruder to recover the secret payload. From equations (1), (2) and (7), it is evident that the Down Scaled Frequency and error values for a grayscale are dependent on the characteristics of the secret image (payload). Therefore it is practically impossible for an intruder to predict a particular payload from DSF or the error matrix of an anonymous payload.

### 3.2 Matrix Embedding

A grayscale cover image $C_i$ is divided into eight bit planes denoted as $C_{ij}$ where $i, j = 0, 1..., 7$. $C_{ij}$ is the $j^{th}$ bit plane of the $i^{th}$ cover image. Embedding process make use of two least significant bit planes of the cover image; let us denote them as $C_{i0}$ and $C_{i1}$. DSF matrix is divided into eight bit planes; let us denote them as $D_i$ where $i= 0, 1, 2....7$. Most significant seven bit planes of the stego images are same as the bit planes of the respective cover images. The least significant bit plane of a stego image is computed as

$$C_{i0} = C_{i1} \oplus D_i, i = 0,1,2,....7 \quad (11)$$

Similarly, error matrix and sign matrix are also embedded into different cover images using equation (12) and (13) respectively.

$$C_{i0} = C_{i1} \oplus E_i, i = 0,1,2,....7 \quad (12)$$

$$C_{00} = C_{01} \oplus S_0 \qquad (13)$$

Here $E_i$ denotes the *ith* bit plane of the error matrix and $S_0$ denotes the only bit plane of the sign matrix.

It is intuitive that only the least significant bit plane of the cover image is modified by replacing the bits with the bits produced by the X-OR operation, denoted as "$\oplus$", between the second least significant bit plane of the cover image and one of the bit planes of the matrices. There are eight bit planes of the DSF matrix, eight bit planes of the error matrix and one bit plane of the sign matrix. To embed all these bit planes seventeen grayscale cover images are required. Diffusion of the payload into different matrices and distribution of matrices over different cover images make it complex enough for the intruder to extract the hidden payload accurately. We use a sample cover image to show the embedding procedure for DSF matrix of the sample payload. Same procedure is used to embed the error as well as sign matrices. Figure 3(a) shows a sample cover image. Eight bit planes of the sample cover image are shown in Figure 3(b)-3(i). The least significant bit plane of the DSF matrix (shown in Figure 2(c)) of the sample payload is shown Figure 3(j). New bit plane (shown in Figure 3(k)) is computed by taking X-OR of the least significant bit plane of the DSF matrix (shown in Figure 2(c)) and second least significant bit plane (shown in Figure 3(h)) of the sample cover image. In the stego image the newly computed bit plane (shown in Figure 3(k)) replaces the least significant bit plane (shown in Figure 3(i)) of the sample cover image. All other bit planes of the stego image are same as the bit planes of the sample cover image. So, one bit plane of the DSF matrix of the sample payload is embedded into the sample cover image. Remaining bit planes of the DSF matrix can be embedded into the cover images using the same procedure.

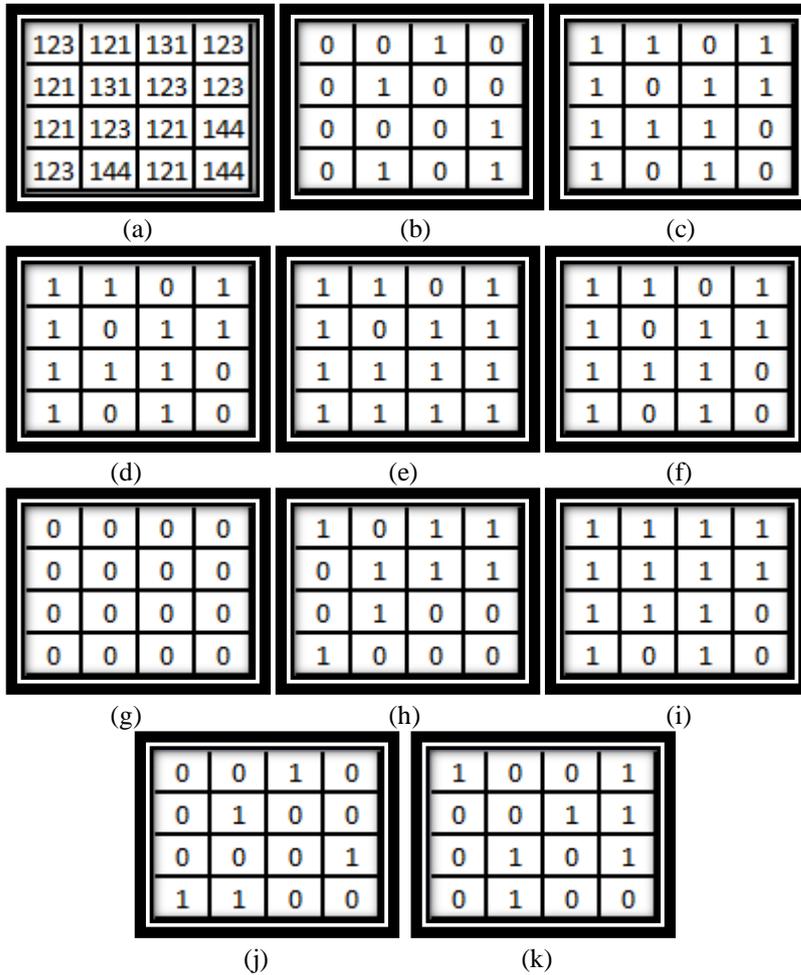

**Fig. 3.** Sample cover image and its bit planes (a)-(i); (a) sample cover image, (b)-(i) bit planes, (b) most significant bit plane, (i) least significant bit plane, (j) least significant bit plane of DSF matrix of sample payload, (k) bit plane obtained by X-ORing (h) and (j).

### 3.3 Payload Extraction

The matrices are extracted from stego images by taking X-OR of second least significant bit plane with the least significant bit plane of the stego image. After the extraction of the DSF matrix, error matrix and sign matrix payload is recomputed as

$$R = \left[ r_{ij} \right]_{m \times n} \text{ where } r_{ij} = \begin{cases} d_{ij} + e_{ij}, & \text{if } s_{ij} = 0 \\ d_{ij} - e_{ij}, & \text{if } s_{ij} = 1 \end{cases} \quad (14)$$

The recovery of payload in the proposed method is computationally efficient as compared to any image steganography scheme employing state of art encryption technique in view of the fact that only addition or subtraction is performed on whether the corresponding value of $s_{ij}$ in the sign matrix is zero or one. Extraction process can retrieve matrices in parallel thereby reducing the computation time two folds.

## 4 Experimental Results

The proposed scheme has been analyzed on satellite images since the difference in pixel intensities of the neighboring pixels is very small. As a result significant noise has to be induced to hide the visual features of a satellite image. Tests have been performed on four satellite images, shown in Fig. 4, to observe the visual distortion introduced in three matrices. Size of each payload tested is 512×512 pixels. Embedding capacity of the scheme is 1 bpp, which is same as LSB replacement steganography scheme [10]. Original payload and corresponding matrices are shown in Fig. 5. Matrices are computed using the mapping parameter $\mu = 256$ and $\eta = 8$. Difference in quality of original payload and the resulting matrices is quite substantial to deceive the intruder into thinking that some kind of noise has crept into the original payload. Fig. 6, Fig. 7 and Fig. 8 show the bit planes of original payload Vrindavan India, corresponding error matrix and frequency matrix respectively. The bit planes of the matrices are visually different from the bit planes of original payload. Distortion introduced in the bit planes of the error and frequency matrices is even more. As the bit planes of the matrices are transmitted through the channel, it is very difficult to predict the original payload from one or two bit planes of the matrices. The extracted payload bears no visual dissimilarity with the original payload. Results show that the proposed scheme does well in concealing the visual features of original payload. Suspicion induced by state of art steganalysis schemes and subsequent retrieval of the hidden payload can adversely affect the secret communication. Hence the visual distortion instigated in the resulting matrices and the corresponding bit planes prevents unauthorized recovery of the secret payload with optimal computation time.

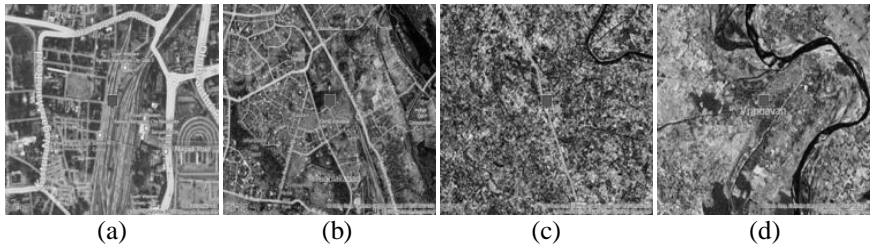

**Fig. 4.** Original payload: (a) Bangalore junction, (b) Okhla India, (c) Patti India, (d) Vrindavan India (Source: Map data ©2012 Google Imagery ©2012 Cnes/Spot Image, DigitalGlobe, GeoEye)

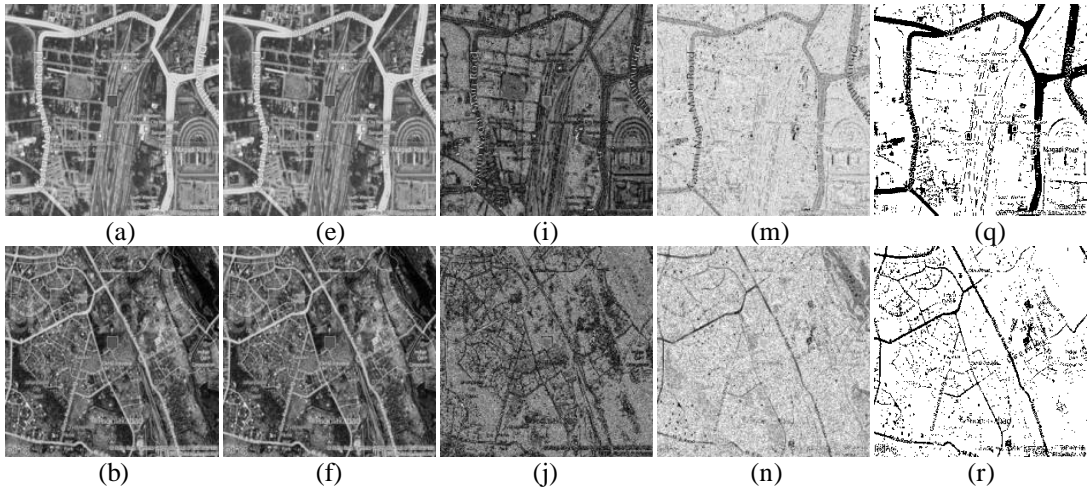

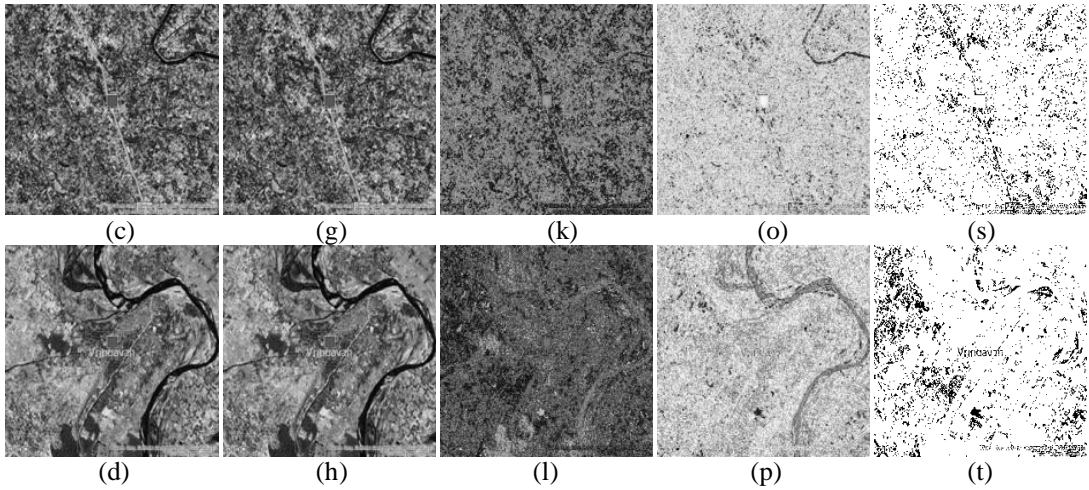

**Fig. 5.** Original payload (a)-(d), recovered payload (e)-(h), error matrix (i)-(l), DSF matrix (m)-(p), sign matrix (q)-(t).

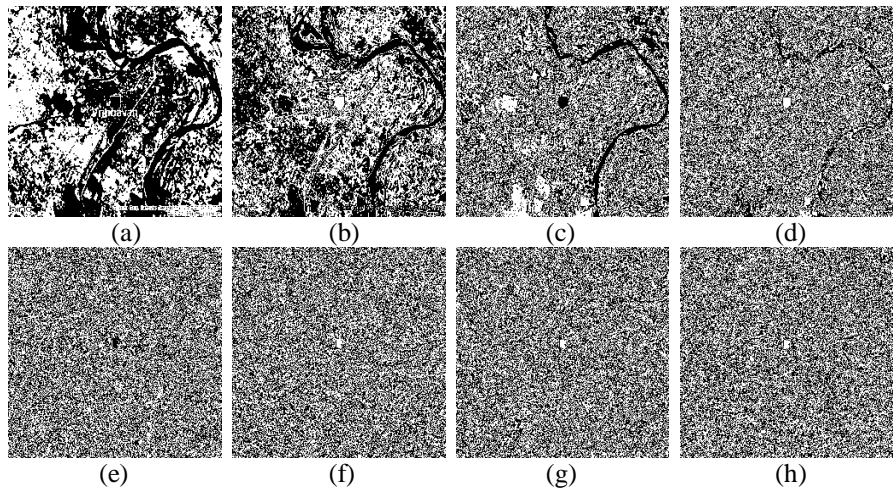

**Fig. 6.** Bit plane of original payload Vrindavan India (a)-(h), most significant bit plane (a), least significant bit plane (h), intermediate bit plane (b)-(g).

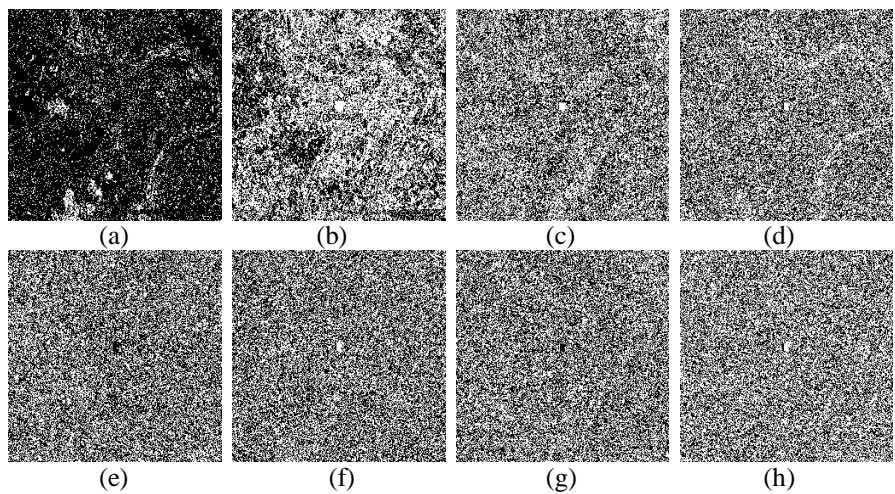

**Fig. 7.** Bit plane of error matrix of original payload Vrindavan India (a)-(h), most significant bit plane (a), least significant bit plane (h), intermediate bit plane (b)-(g).

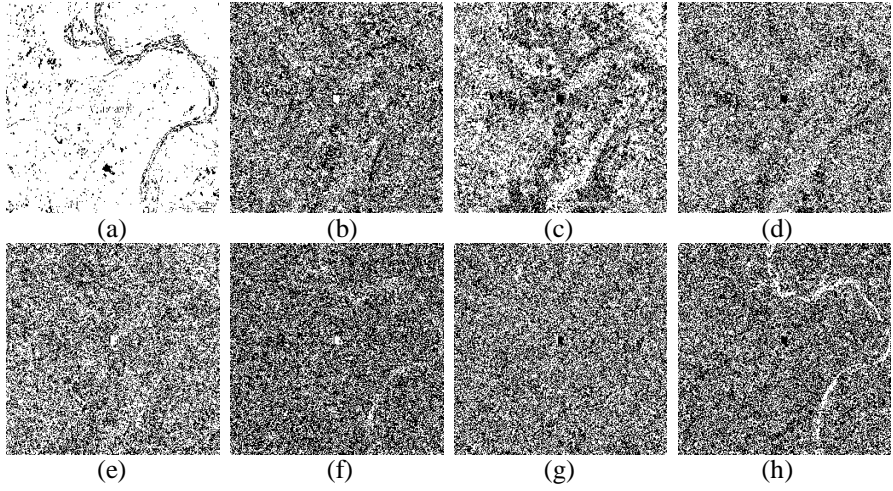

**Fig. 8.** Bit plane of frequency matrix of original payload Vrindavan India (a)-(h), most significant bit plane (a), least significant bit plane (h), intermediate bit plane (b)-(g).

## 4.1 Distortion Strength

Strength of visual distortion in resulting matrices can be measured and analyzed with corresponding peak signal to noise ratio (PSNR) [17]. PSNR is defined as in equation (15). It is the ratio of the square of maximum grayscale intensity to the mean square error (MSE) [18] of the original payload and the corresponding matrix.

$$PSNR = 10\log_{10}\left(\frac{I_{\max}^2}{MSE}\right)(dB) \qquad (15)$$

MSE for DSF matrix is defined as in equation (14).

$$MSE = \frac{1}{MN}\sum_{i=1}^{M}\sum_{j=1}^{N}\left(\left|g_{ij}-d_{ij}\right|\right)^2 \qquad (16)$$

Similarly MSE can be computed for error and sign matrices using equation (16). Table 1 shows PSNR values for four test images and their corresponding matrices. Smaller the PSNR value stronger is the distortion in the matrices. Minimum PSNR for indistinguishable visual distortion is 30dB [3]. Table 1 illustrates that enough noise has been introduced in all three matrices for each payload under consideration to impede the precise interpretation of the satellite imagery.

**Table 2.** PSNR of three matrices for respective payloads

| Original Payload | Error matrix PSNR (dB) | DSF matrix PSNR (dB) | Sign matrix PSNR (dB) |
|---|---|---|---|
| Bangalore Junction | 9.0863 | 7.9396 | 4.1143 |
| Okhla India | 9.7589 | 7.0230 | 3.4600 |
| Patti India | 9.1781 | 6.9692 | 3.6982 |
| Vrindavan India | 9.7201 | 8.5998 | 4.1636 |

## 4.2 Statistical Analysis

We have analyzed the statistical distribution of grayscales in the original payload and the corresponding DSF and error matrices using image histogram. An image histogram graphically illustrates the presence of number of pixels of a particular grayscale in an image [19]. Fig. 9., Shows the histograms of original payloads and their corresponding DSF and error matrices. Histograms for DSF and error matrices are visually distinct from the histogram of the original payload. Different statistical attacks are used to predict the original payload from the encrypted payload. DSF and error matrices are distinctive enough to avoid any statistical attack, which can be used to predict the original payload from the histograms of DSF and error matrices.

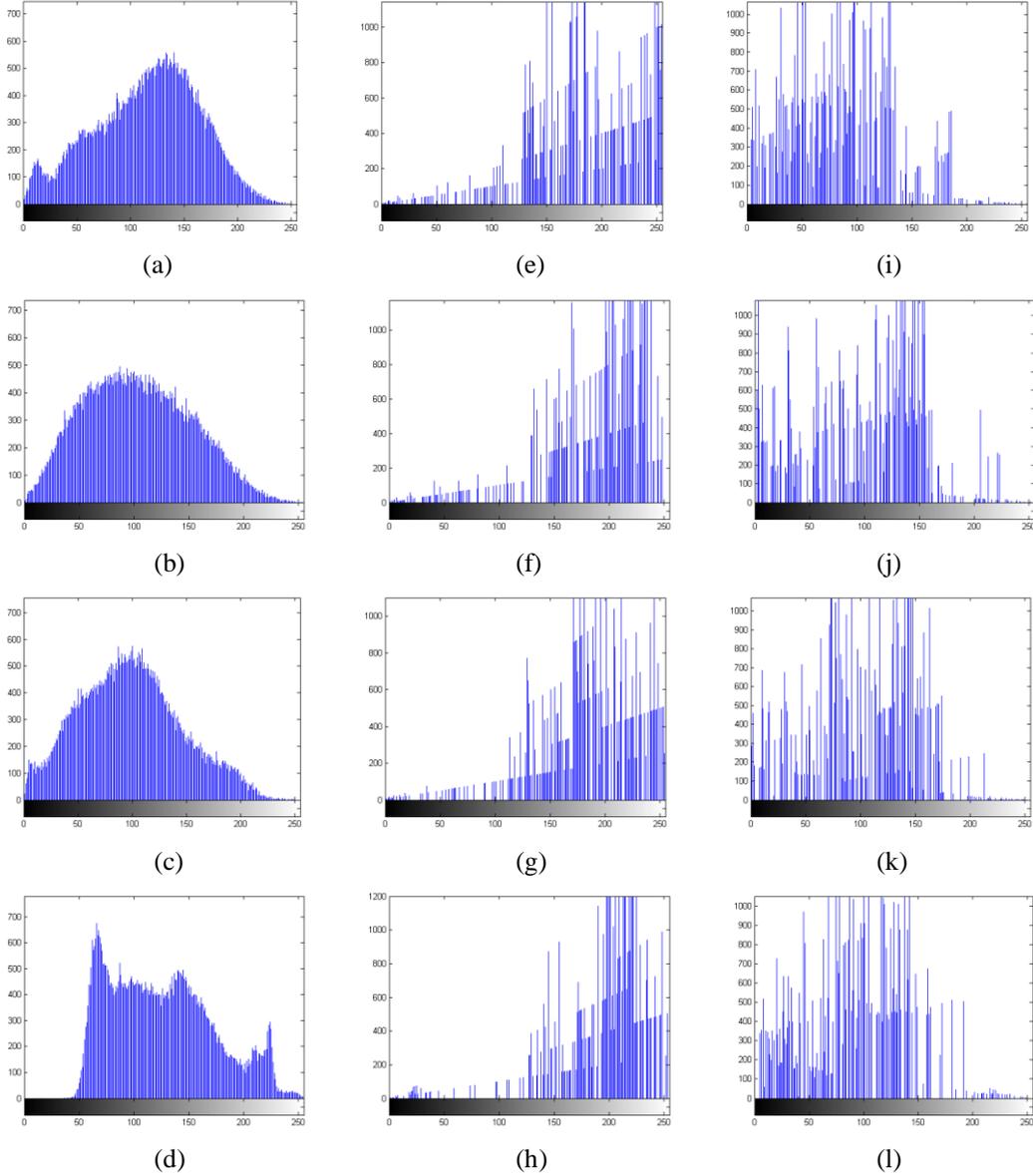

**Fig. 9.** Histogram of original payloads and their respective DSF matrix and error matrix; histograms of original payloads (a)-(d), histogram of Vrindavan India (a), histogram of Patti India (b), histogram of Okhla India (c), histogram of Bangalore junction (d), histogram of corresponding dsf matrix (e)-(h), histogram of corresponding error matrix (i)-(l).

### 4.3 Sensitivity Analysis

We analyze the sensitivity [19] of the proposed scheme with respect to the payload which is distributed over three matrices. We define sensitivity as the strength of linear relationship between the payload and the corresponding matrices. Correlation coefficient is an appropriate objective measure for the strength of linear relationship amongst the payload and the corresponding matrices. The correlation coefficient is defined as

$$r = \frac{\sum_M \sum_N (G_{MN} - \overline{G})(MAT_{MN} - \overline{MAT})}{\sqrt{\left(\sum_M \sum_N (G_{MN} - \overline{G})^2\right)\left(\sum_M \sum_N (MAT_{MN} - \overline{MAT})^2\right)}}$$

(17)

Where $G$ is the payload, $\overline{G}$ is the mean value of the grayscales of the payload, $MAT$ is the corresponding DSF matrix or error matrix or sign matrix and $\overline{MAT}$ is the mean value of the matrix.

If sensitivity is more, then for payloads with slightly different characteristics, the proposed scheme should be able to generate completely different DSF, error and sign matrices. The sensitivity of the proposed scheme is more if there is a weak linear relationship between the payload and a particular matrix. If the value of the correlation coefficient is either between 0 and 0.7 or between 0 and -0.7 the strength of the linear relationship between payload and the corresponding matrix is considered to be weak [20]. Table 3 illustrates correlation coefficients for different payloads and the corresponding matrices. Results show that for different payloads the proposed scheme achieves acceptable correlation coefficients; hence high sensitivity.

**Table 3.** Correlation coeeficients for different payloads and corresponding matrices

| Images | DSF matrix[r] | Error matrix[r] | Sign matrix[r] |
|---|---|---|---|
| Vrindavan India | 0.1175 | -0.4232 | -0.5350 |
| Patti India | -0.2438 | -0.6240 | -0.5436 |
| Okhla India | -0.1722 | -0.5809 | -0.6063 |
| Bangalore Junction | -0.4492 | -0.6066 | -0.6150 |

### 4.4 Computational Complexity

If we consider a payload of size $N \times N$ of 256 grayscales total number of comparisons to compute the frequency of occurrence of each gray level in the payload is at most $256 \times N^2$. The frequency of any gray level can have a maximum value of $N^2$, hence the division with the corresponding divisor which in this case is $level = i + 1 \approx \dfrac{N^2}{\mu}$, where $\mu$ is the mapping parameter, involve at most $\mu$ subtractions which is a constant. For $N^2$ frequency values computational cost of DSF matrix is $\mu \times N^2$.

There are $N^2$ pixel values in the original payload and $N^2$ DSF values in DSF matrix, computational complexity for error matrix computation is at most $N^2$ as there will be $N^2$ subtractions. For sign matrix evaluation the total number of comparisons is at most $N^2$. So, the total computational complexity for matrix decomposition is given by

$$O(256N^2 + \mu N^2 + N^2 + N^2) = O(256N^2 + \mu N^2 + 2N^2) \approx O(N^2)$$
(18)

### 4.5 Time Analysis

The major goal of the proposed payload distribution scheme is to reduce the computation time. We have implemented the scheme and computed the time elapsed in distribution of the payloads to different matrices. The proposed scheme has been implemented using MATLAB 2010 on a system with 2.10GHz-Intel (R) Core™ i3-2310M CPU, 3GB DDR2 RAM and 64-bit windows 7 operating system. The time taken to distribute each payload of different size over three matrices has been shown in Table 4. Table 4 also illustrates the computation time taken by two state-of-art encryption algorithms, AES and DES, to encrypt same set of payloads. Results show that the proposed scheme achieves significant reduction in time for distribution and recovery of the payloads. Especially recovery process shows considerable reduction in computation time.

**Table 4.** Computation time for payload distribution and recovery using the proposed scheme and computation time for encryption and decryption of the payload using state-of-art encryption techniques

| Image | Size | Proposed Method | | DES | | AES | |
|---|---|---|---|---|---|---|---|
| | | Matrix Generation time(s) | Recovery time(s) | Encryption time(s) | Decryption time(s) | Encryption time(s) | Decryption time(s) |
| Vrindavan | $20 \times 20$ | 0.0624 | 0.0035 | 18.2053 | 18.0582 | 33.0098 | 33.3480 |
| Pattiindia | | 0.1560 | 0.0103 | 15.9653 | 15.6285 | 32.7602 | 32.5925 |
| Okhla India | | 0.0624 | 0.0083 | 17.4409 | 17.7351 | 33.0098 | 32.9358 |
| Bangalore Junction | | 0.0624 | 0.0083 | 18.0961 | 18.2639 | 32.1830 | 32.5281 |
| Vrindavan | $30 \times 30$ | 0.0936 | 0.0058 | 40.8723 | 41.0526 | 62.4482 | 32.5724 |
| Pattiindia | | 0.1958 | 0.0063 | 38.1422 | 38.9533 | 63.7742 | 33.9337 |
| Okhla India | | 0.0624 | 0.0091 | 40.9815 | 40.9462 | 62.4170 | 32.3684 |
| Bangalore Junction | | 0.0780 | 0.0085 | 41.3403 | 41.3857 | 62.3546 | 32.4879 |
| Vrindavan | $40 \times 40$ | 0.0936 | 0.0048 | 69.2332 | 69.3827 | 134.5197 | 134.4894 |
| Pattiindia | | 0.1980 | 0.0085 | 69.5608 | 69.7250 | 136.3137 | 136.7465 |
| Okhla India | | 0.0780 | 0.0072 | 74.0849 | 73.9268 | 133.9581 | 133.4813 |
| Bangalore Junction | | 0.0780 | 0.0066 | 70.7777 | 70.4926 | 138.6225 | 138.3654 |
| | | | | | | | |
| Vrindavan | | 0.1092 | 0.0185 | 98.6238 | 98.4729 | 311.6900 | 310.0963 |
| Pattiindia | 50×50 | 0.1996 | 0.0168 | 100.0746 | 100.8364 | 313.0472 | 313.8764 |
| Okhla India | | 0.0936 | 0.0156 | 99.9498 | 100.0463 | 308.6012 | 308.9640 |
| Bangalore Junction | | 0.0936 | 0.0156 | 100.215 | 100.0472 | 304.358 | 308.4875 |
| | | | | | | | |
| Vrindavan | | 0.9360 | 0.0386 | 2105.1 | 2037.3 | 8673.2 | 8427.5 |
| Pattiindia | 250×250 | 0.9644 | 0.0312 | 2109.3 | 2185.6 | 7875.5 | 7993.3 |
| Okhla India | | 0.8580 | 0.0415 | 1904.8 | 2011.4 | 7433.3 | 7493.7 |
| Bangalore Junction | | 0.8268 | 0.0381 | 2012.8 | 2062.7 | 7725.2 | 7738.5 |
| | | | | | | | |
| Vrindavan | | 3.9624 | 0.1674 | 7818.8 | 7493.4 | 33962 | 33854 |
| Pattiindia | 512×512 | 4.0560 | 0.1248 | 7947.5 | 8046.2 | 34288 | 34753 |
| Okhla India | | 3.9468 | 0.1382 | 7483.8 | 7628.5 | 32269 | 32587 |
| Bangalore Junction | | 4.0404 | 0.1936 | 7837.7 | 7883.4 | 33593 | 33897 |

## 5  Conclusions

Extent of visual distortion introduced in the matrices is enough to conceal the payload without adding significant computational complexity to the steganography system. Primary objective of the proposed scheme is to achieve higher level of confidentiality in absence of standard encryption techniques which add up to complexity of the stego system. Experimental results show that the proposed scheme effectively achieve the objective. Even though the proposed scheme has been implemented and analyzed for satellite images, can be used for any grayscale image.